\documentclass{article}

\usepackage[nonatbib, preprint]{neurips_2021}

\usepackage[utf8]{inputenc} 
\usepackage[T1]{fontenc}    
\usepackage{hyperref}       
\usepackage{url}            
\usepackage{booktabs}       
\usepackage{amsfonts}       
\usepackage{nicefrac}       
\usepackage{microtype}      
\usepackage{xcolor}         
\usepackage{tikz}
\usepackage{pgfplots}
\usepackage{multirow}
\usepackage{graphicx}
\usetikzlibrary{pgfplots.groupplots, external, positioning }

\usepackage{amsmath}
\usepackage{algorithm}
\usepackage{algpseudocode}
\usepackage{bm}
\pgfplotsset{compat=1.5}

\title{A fully-differentiable compressible high-order computational fluid dynamics solver}

\author{%
  Deniz A. Bezgin\thanks{D.A.B. and A.B.B. contributed equally to this work.} \\
  Chair of Aerodynamics and Fluid Mechanics \\
  Technical University of Munich \\
  D-85748 Garching bei Muenchen \\
  \texttt{deniz.bezgin@tum.de} \\
  \And
  Aaron B. Buhendwa\footnotemark[1] \\
  Chair of Aerodynamics and Fluid Mechanics \\
  Technical University of Munich \\
  Garching bei Muenchen \\
  \texttt{aaron.buhendwa@tum.de} \\
  \AND
  Nikolaus A. Adams \\
  Chair of Aerodynamics and Fluid Mechanics \\
  Technical University of Munich \\
  Garching bei Muenchen \\
  \texttt{nikolaus.adams@tum.de} \\
}

\begin{document}

\maketitle

\begin{abstract}
  Fluid flows are omnipresent in nature and engineering disciplines.
  The reliable computation of fluids has been a long-lasting challenge due to nonlinear interactions over multiple spatio-temporal scales.
  The compressible Navier-Stokes equations govern compressible flows and allow for complex phenomena like turbulence and shocks.
  Despite tremendous progress in hardware and software, capturing the smallest length-scales in fluid flows still introduces prohibitive computational cost for real-life applications. 
  We are currently witnessing a paradigm shift towards machine learning supported design of numerical schemes as a means to tackle aforementioned problem.
  While prior work has explored differentiable algorithms for one- or two-dimensional incompressible fluid flows, we present a fully-differentiable three-dimensional framework for the computation of compressible fluid flows using high-order state-of-the-art numerical methods.
  Firstly, we demonstrate the efficiency of our solver by computing classical two- and three-dimensional test cases, including strong shocks and transition to turbulence.  
  Secondly, and more importantly, our framework allows for end-to-end optimization to improve existing numerical schemes inside computational fluid dynamics algorithms.
  In particular, we are using neural networks to substitute a conventional numerical flux function. 
\end{abstract}

\section{Intro}
\label{sec:Intro}

Fluid flows are omnipresent in nature and engineering disciplines with applications reaching from the design of combustion-engines and windturbines,
the simulation of climate and weather forecasts, to magnetohydrodynamics in plasma physics.
The accurate simulation and the development of high-fidelity models for such flows has been a long-standing challenge in computational fluid mechanics.

In the past few years machine learning has been extensively used in engineering disciplines and physical sciences alike.
ML methods have the capabilities to learn complex relations from data and enable novel modeling strategies as well as new avenues of postprocessing.
Amongst others, the intersection of machine learning and fluid mechanics has experienced renewed interest \cite{Brunton2020a}.
A multitude of research has put forward different ways of using ML in scientific applications.
These methods can be categorized according to different criteria.
One important distinction is the ratio of physical prior-knowledge to "pure ML" that is included in model and training.
While pure ML schemes have the advantage of being efficient and quickly implemented, they do not offer guarantees on performance (e.g. convergence or stability).
Since they do not include the underlying physics, it is often diffcult to enforce constraints such as symmetry or conversation of energy.
In contrast, established numerical methods often come with guarantees on convergence and error bounds, and a plethora of research has investigated the hybridization of ML and classical numerics.

A second major distinction of ML models can be made according to on- and offline training.
Up unitl now, ML models have been typically optimized offline, i.e. outside of a simulator.
Upon proper training, they are then plugged in the solver for evaluation of down-stream tasks.
Examples include training of explicit subgrid scale models in large eddy simulations \cite{Beck2019}, interface reconstruction in multiphase flows \cite{Patel2019a,Buhendwa2021b}, and cell-face reconstruction in shock-capturing schemes \cite{Stevens2020c,Bezgin2021b}.
However, the successful development of powerful general-purpose automatic-differentiation frameworks, such as Tensorflow \cite{Abadi}, Pytorch \cite{Paszke2019}, and JAX \cite{jax2018github}
has enabled online training of ML models.
Models are optimized end-to-end within a differentiable simulator or programm.
This has two significant advantages: the ML model observes the dynamics of the underlying physics and sees its own outputs as inputs.
In this fashion, Bar-Sinai et al. \cite{Bar-Sinai2019} have proposed data-driven discretizations for one- and two-dimensional PDEs while Bezgin et al. \cite{Bezgin2021a} have trained a subgrid-scale closure for nonclassical shocks.
More recently, Schoenholz and Cubuk have put forward JAX-MD \cite{Schoenholz}, a differentiable software package for molecular dynamics. 
An exciting work in the direction of fluid mechanics is JAX-CFD \cite{Kochkove2101784118} which is an ML-accelerated fluid dynamics simulator for incompressible two-dimensional flows.
JAX-CFD has been successfully used to learn data-driven turbulence models for two-dimensional turbulent flows.

Inspired by these previous works, we introduce a fully differentiable high-order computational fluid dynamics solver for the compressible Navier-Stokes equations.
As previous works have focused on one- or two-dimensional systems, our framework allows for the simulation of the three-dimensional compressible Navier-Stokes equations. 
Compressible flows are challenging as they feature not only chaotic turbulent behavior but may also be shock-dominated.
Suitable numerical methods, therefore, have to be capable of resolving small-scale flow features and of shock capturing.
These at times contradictory requirements have been a long-lasting challenge and require novel solution which we want to explore with differentiable simulators.  
Therefore, our numerical solver is complelety written in JAX - a language that supports reverse-mode automatic differentiation.
This allows us end-to-end training of ML models like neural-networks within the solver.

In the following we detail physical and numerical aspects of the fluid solver alongside important design and programming aspects.
We demonstrate the efficiency of the forward-pass through our solver.
Finally, we illustrate how our framework can be used for exploration of data-driven numerical schemes. 

\section{CFD: Physical and numerical models}

\subsection{Governing equation}




A general differential conservation law for the state vector $\bm{U}$ can be written in symbolic notation as 

\begin{align}
    \frac{\partial \bm{U}}{\partial t} = - \nabla\bm{F}\left(\bm{U}\right) + S(\bm{U}),
    \label{eq:ConservationLaw}
\end{align}

where $\bm{F}(\bm{U})$ and $S(\bm{U})$ are the flux vector and the source term vector \cite{Toro2009a}.
For the compressible Navier-Stokes equations, the state vector of conservative variables $\bm{U} = \left(\rho, \rho u, \rho v, \rho w, E\right)^T$. 
Here, $\rho$ is the density, $u,v,w$ are the velocity components in the spatial directions $x,y,z$, respectively, and $E$ is the total energy.
A different representation of the state is given by the vector of primitive variables $\bm{W} = \left(\rho, u, v, w, p\right)^T$, with the pressure $p$.
The total energy is given by $E = \rho e + \frac{1}{2} \rho (u^2 + v^2 + w^2)$, with the internal energy per unit mass $e$.
An equation of state is needed to close the system of equations.
In this work, we choose the equation of state for an ideal gas, $e = p / \left( (\gamma - 1) \rho \right)$ with the ratio of specific heats $\gamma$.
We set $\gamma = 1.4$ if not stated otherwise.\\

For the Navier-Stokes equations, the source term vector includes viscous forces, heat conduction, and gravitational forces, amongst others.
For $S(\bm{U}) = 0$, we recover the Euler equations which are a set of hyperbolic differential equations \cite{Toro2009a}.

\subsection{Architecture}

We solve the compressible Navier-Stokes equations using finite volumes on a Cartesian mesh with cubic cells in 3D.
The simplified compute loop of our code is depicted in algorithm \ref{alg:ComputeLoop}. 
Starting from the initial vector of conservatives $\bm{U}_0 = \bm{U}(t_0)$ at time $t_0$, integration steps are performed until the final time $t_{end}$ is reached.
A single integration step consists mainly of the right hand side evaluation of Eq. \eqref{eq:ConservationLaw} and subsequent integration of the vector of conservatives. 
The majority of compute time is spend in $\mathtt{compute\_right\_hand\_side}(\bm{U}^n)$.
Here, the general procedure consists of the left and right sided cell-face reconstruction $\bm{U}_L$, $\bm{U}_R$ and subsequent evaluation of the cell-face Riemann problem, yielding the
cell-face flux $\bm{F}^n$ (compare algorithm \ref{alg:ComputeRightHandSide}).

In literature, there exists a wide variety of different time integrators, cell-face reconstruction schemes, and approximate Riemann solvers \cite{Toro2009a}.
We use a modular, object-oriented programming framework which allows for convenient exchange of submodules, i.e. time integration schemes, cell-face reconstruction schemes, and Riemann solvers. 
Before running the simulator, the user specifies the numerical setup. 
We focus on state-of-the-art high-order shock-capturing methods.
Explicit total-variation-diminishing (TVD) Runge-Kutta integrators up to order 3 and the classical Runge-Kutta 4 schemes are available \cite{Shu1987,Gottlieb1998a}.
For the cell-face reconstruction, we provide weighted essentially non-oscillatory (WENO) schemes and its variants up to order 9 \cite{Jiang1996,Borges2008a}.
Choices for approximate Riemann solvers include the Rusanov-flux \cite{Toro2009a}, HLL, and HLLC solver.

Our solver is completely written in JAX \cite{jax2018github} which is an updated version of Autograd and XLA.
JAX comes with a fully-differentiable version of the popular NumPy package \cite{Harris2020} called JAX NumPy.
In our implementation we make heavy use of JAX NumPy.
The primitive and conservative variable vectors are stored as arrays of size $(5, n_x + 2 n_h, n_y + 2 n_h, n_z + 2 n_h)$, where $(n_x, n_y, n_z)$ is the resolution in the three spatial dimensions and $n_h$ are the number of halo cells.
Halo cells are required to impose boundary conditions.
The algorithm naturally degenerates for one- or two-dimensional problems by using a single cell in the excess dimensions.

Atop of the pure numerical computing functionality, JAX provides further function transformations.
All compute-intensive routines in our forward-pass are just-in-time compiled via the $\mathtt{jit}$ function.
In order for JAX transformations to work properly, these routines must be expressed as pure functions, i.e. variables that change during the simulation must be passed as function arguments.
When using the fluid solver for training and evaluation of ML models, we need tools for differentiation and batch evaluation.
For automatic differentiation we use $\mathtt{grad}$ and $\mathtt{value\_and\_grad}$.
Evaluating a batch of simulations in parallel is ensured by the vectorization function $\mathtt{vmap}$.
Neural networks are built via Haiku \cite{haiku2020github} and trained via Optax \cite{optax2020github}.

\begin{algorithm}[t]
    \caption{Simplified compute loop for Euler method.}
    \label{alg:ComputeLoop}
    \begin{algorithmic}
    \Require $\bm{U}^n, t_{end}$
    \While{$t < t_{end}$}
    \State $ \Delta t \gets \mathtt{compute\_time\_step}(\bm{U}^n)$
    \State $\bm{U}_{rhs}^n \gets \mathtt{compute\_right\_hand\_side}(\bm{U}^n)$
    \State $\bm{U}^n \gets \mathtt{integrate}(\bm{U}^n, \bm{U}^n_{rhs}, \Delta t)$
    \State $\bm{U}^n \gets \mathtt{fill\_boundaries}(\bm{U}^n)$
    \State $t \gets t + \Delta t$
    \EndWhile
\end{algorithmic}
\end{algorithm}

\begin{algorithm}[t]
    \caption{Simplified algorithm to compute the right hand side.}
    \label{alg:ComputeRightHandSide}
    \begin{algorithmic}
    \Require $\bm{U}^n$
    \State $\bm{U}_L^n, \bm{U}_R^n  \gets \mathtt{reconstruct\_cell\_face\_values}(\bm{U}^n)$
    \State $\bm{W}_L^n, \bm{W}_R^n \gets \mathtt{compute\_primitives\_from\_conservatives}(\bm{U}^n)$
    \State $\bm{F}^n \gets \mathtt{solve\_riemann\_problem}(\bm{U}_L^n, \bm{U}_R^n, \bm{W}_L^n, \bm{W}_R^n)$
    \State $\bm{U}_{rhs}^n \gets \bm{F}^n$
\end{algorithmic}
\end{algorithm}

\section{Results}

\subsection{Forward-pass}

In this section, classical test-cases for fluid flows exhibiting strong shocks and transition to turbulence are shown. The numerical setup consists
of a WENO5-JS \cite{Jiang1996} cell-face reconstruction combined with a HLLC \cite{Toro2009a} Riemann solver and a TVD-RK3 \cite{Gottlieb1998a} integration scheme.
The initial and boundary conditions for the following cases are detailed in the appendix \ref{IC_RTI}, \ref{IC_TGV}, \ref{IC_DMR}

\subsubsection{Rayleigh-Taylor instability}

The (inviscid) Rayleigh-Taylor instability is a gravity driven instability that occurs at the interface between two fluids of different densities when the lighter fluid is pushing
the heavier fluid \cite{Sharp1984}. In Figure \ref{rti}, instantaneous density contours for various resolutions are depicted. The use of high-order methods tremendously decreases
numerical dissipation. Therefore, floating point errors are no longer hidden by numerical dissipation and result in symmetry-breaking behavior \cite{Fleischmann2019a}. This effect
can be seen for the resolutions $512\times2048$ and $1024\times4096$.

\begin{figure}[b]
    \centering
    \begin{tikzpicture}
        \node[inner sep=0cm] (a1) at (0.0,0) {\includegraphics[scale=0.35]{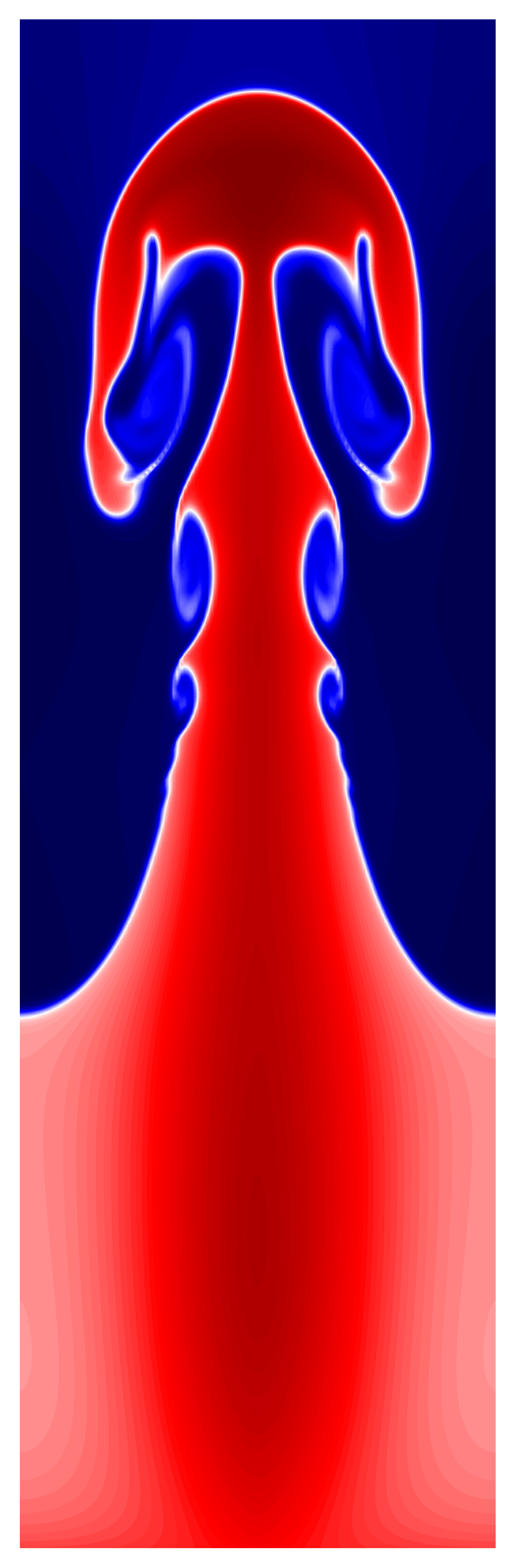}};
        \node[inner sep=0cm, right = 0.5 of a1] (b1) {\includegraphics[scale=0.35]{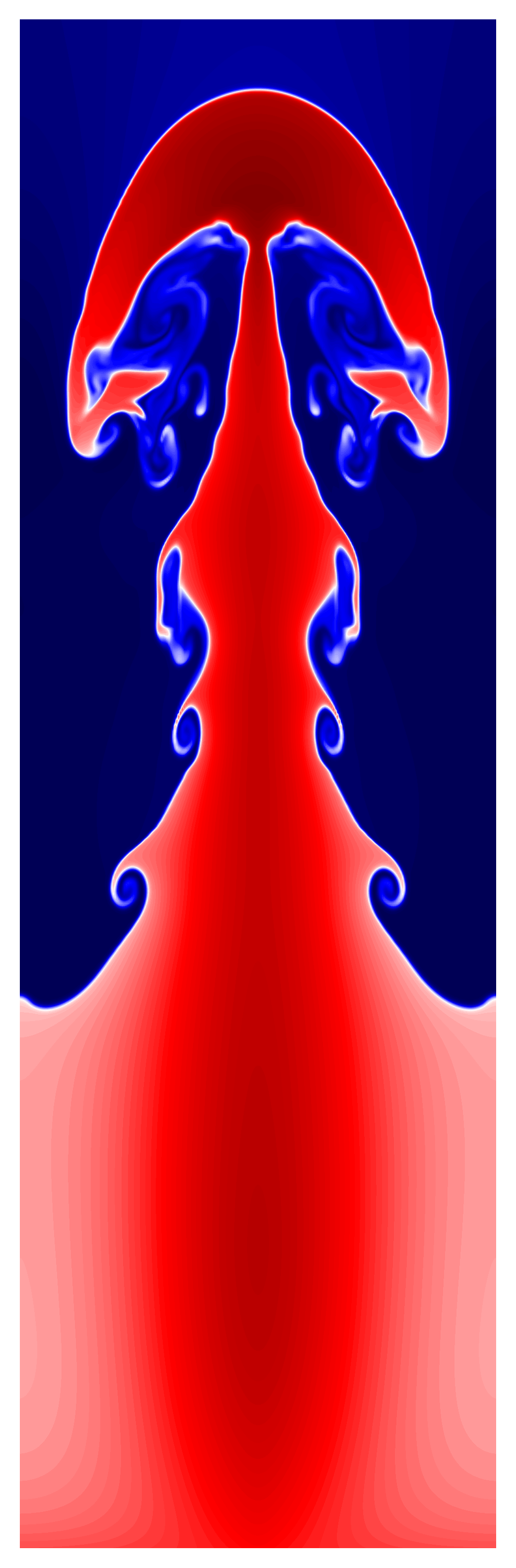}};
        \node[inner sep=0cm, right = 0.5 of b1] (c1) {\includegraphics[scale=0.35]{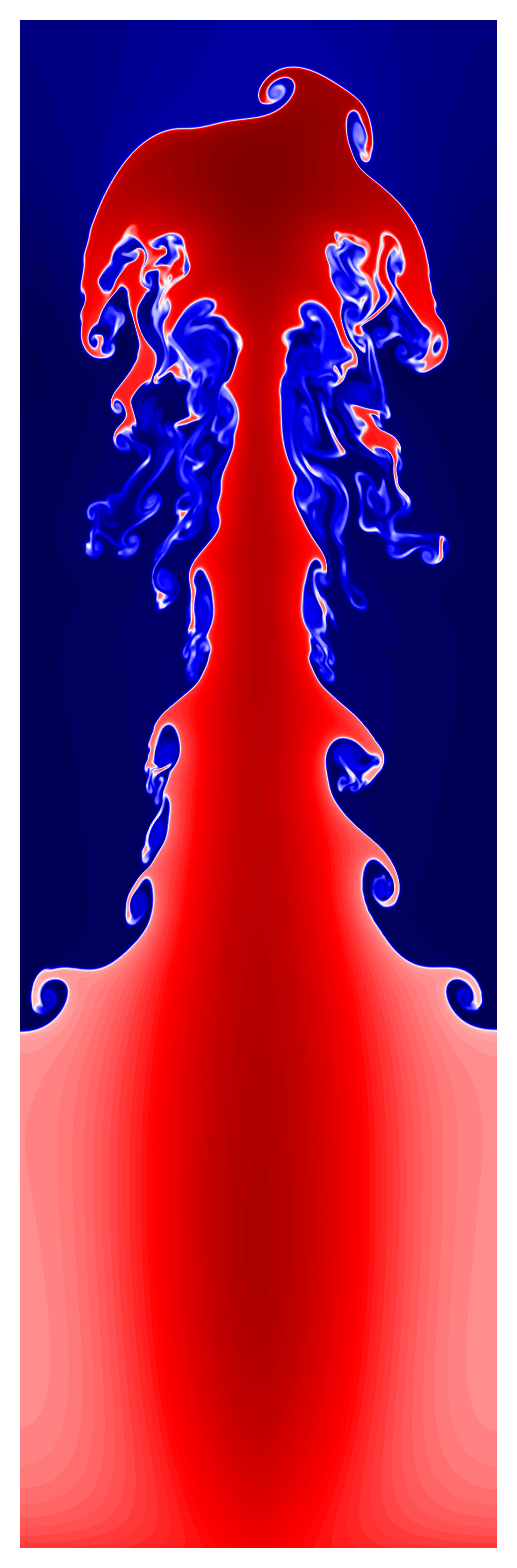}};
        \node[inner sep=0cm, right = 0.5 of c1] (d1) {\includegraphics[scale=0.35]{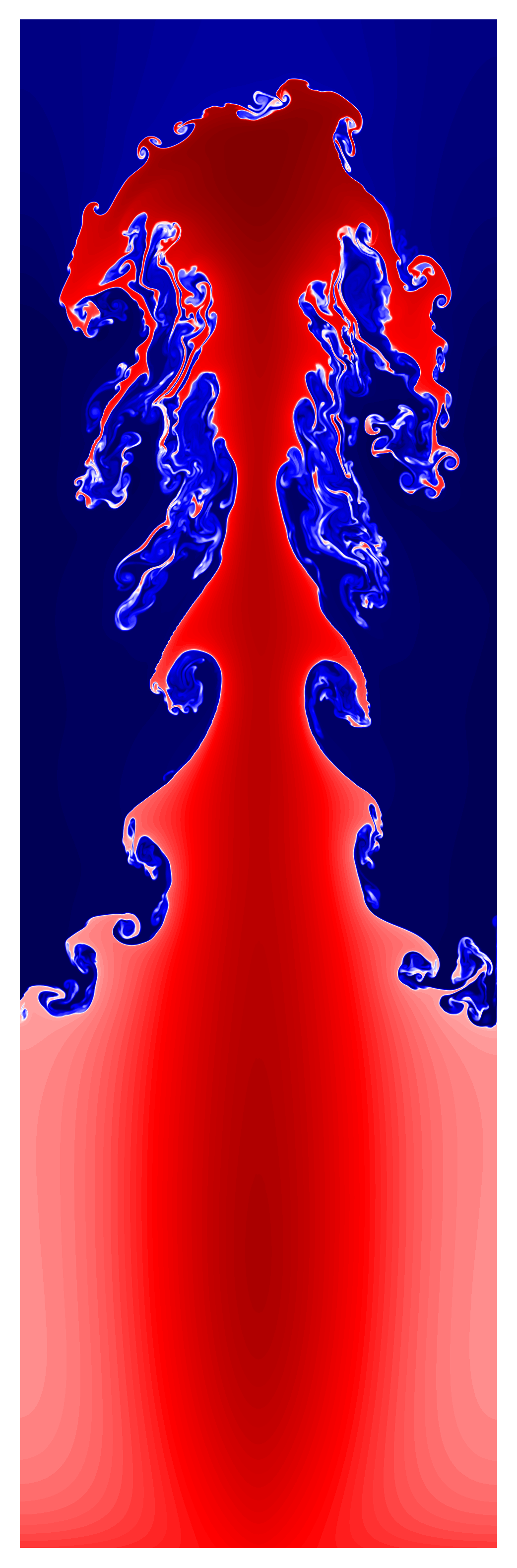}};
        \node[above = 0.0 of a1] {$128\times512$};
        \node[above = 0.0 of b1] {$256\times1024$};
        \node[above = 0.0 of c1] {$512\times2048$};
        \node[above = 0.0 of d1] {$1024\times4096$};
    \end{tikzpicture}
    \label{rti}
    \caption{Instantaneous density contours of the Rayleigh-Taylor instability for various resolutions.}
\end{figure}
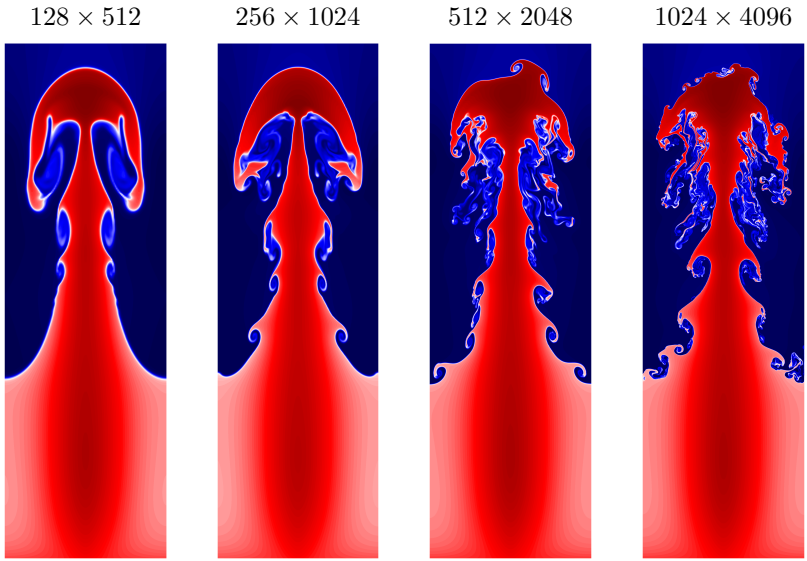

\begin{figure}[t]
    \centering
    \begin{tikzpicture}


\begin{axis}[
width = 7cm,
height = 6cm,
tick align=outside,
tick pos=left,
xmin=-0.498076923076923, xmax=10.4596153846154,
xtick style={color=black},
ymin=-0.000819457758289875, ymax=0.0163470258506987,
ytick style={color=black},
xlabel = {$t$},
ylabel = {$-\frac{\partial K}{\partial t}$},
yticklabels = {0.0, 0.5, 1.0, 1.5},
ytick = {0.00, 0.005, 0.01, 0.015},
legend style={at={(0,1)}, anchor = north west}
]
\addplot [blue]
table {%
0 -3.017283465691e-06
0.1013 -3.40278487011612e-06
0.2001 -1.24196687437418e-05
0.3015 -1.10011447890516e-05
0.4003 -1.98544817852079e-05
0.5016 -1.42744104359984e-05
0.6004 -1.92614172544115e-05
0.7017 -1.23261971909736e-05
0.8005 -1.0680807450776e-05
0.9018 -1.23947132865485e-06
1.0006 7.73748054834206e-06
1.102 1.89860508903e-05
1.2007 3.9853920904914e-05
1.302 5.19602696506196e-05
1.4007 8.58439816937104e-05
1.502 0.000105433715709748
1.6006 0.000150654363958999
1.7017 0.000185366718863293
1.8003 0.000241752847919978
1.9013 0.000295478307151766
2.0022 0.00036780306238019
2.1006 0.000435465535505859
2.2014 0.000526501761055638
2.3021 0.000615542265295578
2.4003 0.000724822276963308
2.501 0.000843629549305095
2.6016 0.000981287726939517
2.7021 0.00113014043434351
2.8002 0.00132239766549669
2.9007 0.00154017962447719
3.0011 0.00182365227252181
3.1016 0.00216958535829185
3.202 0.00257809392119728
3.3025 0.00306300117521571
3.4004 0.0036063489043259
3.5008 0.00422731397376572
3.6012 0.00486387353534727
3.7015 0.00554937917151355
3.8019 0.00618790810983263
3.9022 0.00684831986864114
4.0025 0.00747826268549127
4.1002 0.0080664336964767
4.2004 0.00862490237913855
4.3006 0.00915536648077173
4.4006 0.00963276941507292
4.5005 0.00994216490923431
4.6004 0.0102577953577432
4.7002 0.0105136760672301
4.8 0.0107244065455561
4.9023 0.0109509688167961
5.0021 0.0112389415844223
5.1019 0.0114733797755225
5.2016 0.0116282098830973
5.3014 0.0116889330439709
5.4012 0.0117133575779127
5.501 0.0117193404839999
5.6009 0.0117386020536885
5.7007 0.0117611791538461
5.8006 0.0118108566686557
5.9006 0.0119300549006986
6.0007 0.0120389479847037
6.1008 0.0121434340674247
6.201 0.0122474878353933
6.3013 0.0123819022415439
6.4017 0.0125525150736725
6.5021 0.0126819673270366
6.6 0.0126664184669647
6.7005 0.0125647153971725
6.8011 0.0125169058855499
6.9017 0.0124568505475591
7.0023 0.0122695503421409
7.1005 0.0120504635448753
7.2012 0.0118513175877509
7.3019 0.0116615157789075
7.4001 0.0114677158885904
7.5008 0.0113455135345986
7.6014 0.0112939679201663
7.702 0.0111551063351553
7.8002 0.0108983958568756
7.9009 0.0107025212404501
8.0016 0.0104560898192094
8.1025 0.0101335799962903
8.2009 0.00977310017591638
8.3019 0.00943293723250362
8.4003 0.00908467487624096
8.5012 0.00882841698305383
8.6022 0.00860282410165093
8.7006 0.00839425585664286
8.8016 0.00822076493551296
8.9002 0.00807453972324168
9.0014 0.00798339850971432
9.1001 0.00791390336319589
9.2014 0.00781415118969976
9.3002 0.00767991368861141
9.4016 0.00755762155104218
9.5005 0.00741563628042411
9.602 0.0072773961059073
9.701 0.00716737561714049
9.8025 0.00703042077526854
9.9015 0.00687679380713049
};
\addlegendentry{$64^3$}
\addplot [green]
table {%
0 -5.77922428777213e-06
0.1001 -9.57726654982061e-06
0.2001 -2.14336345117582e-05
0.3002 -2.0890869578139e-05
0.4003 -3.34810280544495e-05
0.5003 -3.37741097870543e-05
0.6004 -4.09515662291859e-05
0.7004 -3.9523836676574e-05
0.8005 -4.66168224559933e-05
0.9006 -4.24706844967537e-05
1.0006 -4.26791920236548e-05
1.1007 -4.38135912642079e-05
1.2007 -3.54714856989902e-05
1.3007 -3.71133631113518e-05
1.4007 -2.53435714564971e-05
1.5006 -2.78718023835786e-05
1.6005 -1.00850217666416e-05
1.7004 -1.22268639682394e-05
1.8002 3.56223568272292e-06
1.9012 8.9600992907018e-06
2.0008 2.37212978262299e-05
2.1004 2.74150606312744e-05
2.2011 4.85651905064643e-05
2.3006 5.37199095611362e-05
2.4012 6.94522371172087e-05
2.5004 8.46037009333304e-05
2.6009 0.000101130497115381
2.7 0.000114317875843163
2.8004 0.000144017410642745
2.9007 0.000163894946343571
3.0009 0.000203395943632466
3.1011 0.000252977916196222
3.2001 0.000323720532327121
3.3003 0.000433769404030287
3.4004 0.000564129385724576
3.5006 0.000745921635483347
3.6007 0.000966365961272434
3.7009 0.00124641956808685
3.801 0.00158564502305599
3.9009 0.0020076988375233
4.0006 0.00247064486096567
4.1001 0.00293204318541245
4.2007 0.00333076239770274
4.3012 0.00367533921841327
4.4003 0.00408646482879299
4.5004 0.00451672729755064
4.6004 0.00491334100064802
4.7003 0.00521535399406152
4.8003 0.00542938192671279
4.9001 0.00552467281911754
5.0011 0.0056114951735242
5.1009 0.00563880436005167
5.2007 0.00568093722497364
5.3005 0.00574021830155575
5.4003 0.00583097681194437
5.5001 0.00601092216709984
5.6012 0.00629675853437994
5.7009 0.00665005282995713
5.8005 0.00700593788962406
5.9 0.00738288006903813
6.0008 0.00771573422512991
6.1004 0.00809181567519756
6.2001 0.00847978393406865
6.3001 0.00883052574333757
6.4004 0.00902118704852562
6.501 0.00912779735129493
6.6007 0.009253758968546
6.7002 0.00949512068428628
6.8008 0.00994425907435996
6.9 0.0103616602925485
7.0004 0.0106872479328664
7.1006 0.0109020046426313
7.2006 0.0110704931813143
7.3005 0.0110577877070099
7.4004 0.0110271761205523
7.5003 0.011165861787692
7.6003 0.0116046864593036
7.7003 0.0122005290267458
7.8004 0.012849761991728
7.9007 0.0133676014959744
8.0001 0.0135150853473274
8.1006 0.0136040100164592
8.2007 0.0134354690783967
8.3009 0.0131279626985227
8.4011 0.0128320750452817
8.5005 0.0126556872611277
8.6001 0.0125084611599165
8.7 0.0120785774335302
8.8001 0.0116550808384734
8.9003 0.0114319471192443
9.0004 0.011330503117774
9.1006 0.011313954549213
9.2008 0.0113315072938743
9.3009 0.0112926467489041
9.401 0.011265703423991
9.5012 0.0112036402085575
9.6002 0.0109960253768413
9.7007 0.0107503067306846
9.8012 0.0104645291882097
9.9006 0.0102920321586145
};
\addlegendentry{$128^3$}
\addplot [red]
table {%
0 -7.76559121480307e-06
0.1001 -1.16386097584464e-05
0.2001 -2.28246125801388e-05
0.3002 -2.17083120130595e-05
0.4003 -3.55001351687312e-05
0.5003 -3.70769026852585e-05
0.6004 -4.34112443015167e-05
0.7004 -4.31053549655867e-05
0.8005 -5.23759712035786e-05
0.9006 -4.79174377221448e-05
1.0006 -5.03976735779629e-05
1.1007 -5.38880930062157e-05
1.2007 -4.66791160277925e-05
1.3007 -4.93915840679704e-05
1.4007 -4.11620140597216e-05
1.5006 -4.5007788574381e-05
1.6005 -2.90162346119085e-05
1.7004 -3.51191370807281e-05
1.8002 -2.1350962882732e-05
1.9012 -1.96965131738081e-05
2.0008 -9.43217202134025e-06
2.1004 -1.15493057584138e-05
2.2011 5.46642384094198e-06
2.3006 2.84402906046663e-06
2.4012 1.06865135667011e-05
2.5004 1.73235771929835e-05
2.6009 2.21538741536517e-05
2.7 2.0887910396193e-05
2.8004 2.92397579528673e-05
2.9007 2.31921428131083e-05
3.0009 2.96514545282946e-05
3.1011 3.33990602738956e-05
3.2001 3.43836168264268e-05
3.3003 4.5973000480574e-05
3.4004 5.88781374315297e-05
3.5006 9.480331903488e-05
3.6007 0.000144254295107031
3.7009 0.000225446352896059
3.801 0.000350313717371491
3.9009 0.000530160654543112
4.0006 0.000738601626816612
4.1001 0.000931661663110998
4.2007 0.00114260181318401
4.3012 0.00133977061582541
4.4003 0.00155690047833923
4.5004 0.00177116143853809
4.6004 0.00200850458032088
4.7003 0.00216416252359978
4.8003 0.00231697927174848
4.9001 0.00241019741433968
5.0011 0.00260951732380453
5.1009 0.00284195069244352
5.2007 0.00304738652335007
5.3005 0.00316113708590944
5.4003 0.00316720530520543
5.5001 0.00313074644877772
5.6012 0.00321829785427804
5.7009 0.00344786527284189
5.8005 0.00373257090632809
5.9 0.00398952814170914
6.0008 0.00439639121146754
6.1004 0.00497506651153346
6.2001 0.00554779578049615
6.3001 0.00604975097258137
6.4004 0.00622767427717164
6.501 0.00611086686430084
6.6007 0.00594414267215109
6.7002 0.00594880497315088
6.8008 0.00622287066019327
6.9 0.00661857157559248
7.0004 0.00709094782934762
7.1006 0.00756286483302923
7.2006 0.00813866168660958
7.3005 0.00847754437000196
7.4004 0.00893622376945183
7.5003 0.00940862282954816
7.6003 0.00984953723435303
7.7003 0.0104379453985661
7.8004 0.0110950109452013
7.9007 0.011817723721534
8.0001 0.0123235565238466
8.1006 0.0130935165196574
8.2007 0.0137218505090191
8.3009 0.0140087563261909
8.4011 0.0142566542448142
8.5005 0.0138298275536804
8.6001 0.0135045995785243
8.7 0.0133677595242615
8.8001 0.0131408475658732
8.9003 0.0129679961128766
9.0004 0.0127219147344735
9.1006 0.0124491614719505
9.2008 0.0122397301465128
9.3009 0.0121876517256811
9.401 0.0122284598067905
9.5012 0.0123887020630302
9.6002 0.0121024948343362
9.7007 0.0120969291547161
9.8012 0.0122724611226075
9.9006 0.012076640426562
};
\addlegendentry{$256^3$}

\addplot [semithick, black, mark=*, mark size=1, only marks]
table {%
0.115384615384615 0.000154738878143132
0.48076923076923 0.000154738878143132
1 0.000185686653771758
1.5 0.000216634429400386
2.01923076923077 0.000216634429400386
2.51923076923077 0.000309477756286267
3 0.000371373307543518
3.51923076923077 0.000464216634429398
4.01923076923077 0.000742746615087039
4.5 0.0012688588007737
5 0.0018568665377176
5.5 0.00238297872340425
6 0.00297098646034816
6.5 0.00445647969052224
7 0.00597292069632495
7.5 0.00761315280464217
8 0.00987234042553191
8.5 0.0146382978723404
8.6923076923077 0.0155667311411992
9 0.0152263056092843
9.5 0.0135860735009671
9.96153846153846 0.0126885880077369
};
\addlegendentry{DNS}
\end{axis}


    \node at (9,2) {
    \renewcommand{\arraystretch}{1.5}
    \begin{tabular}[]{c c c}
        N&$T_1 [s]$&$T_2 [s]$ \\
        \hline
        $64^3$&1.2e-07&1.41e+02\\
        $128^3$&1.3e-07&2.22e+03\\
        $256^3$&1.5e-07&4.09e+04\\
    \end{tabular}

    };
    \end{tikzpicture}
    \caption{(Left) Rate of turbulence kinetic energy dissipation $\frac{\partial K}{\partial t}$ with $K=\frac{1}{2}\left(\bar{u}^2+\bar{v}^2+\bar{w}^2\right)$ for Taylor-Green vortex.
            The DNS data is taken from \cite{Brachet1983}.
            (Right) Average wall clock time per cell per time step $T_1$ and wall clock time for entire simulation $T_2$ on a NVIDIA Quadro P5000.}
    \label{tgv}
\end{figure}
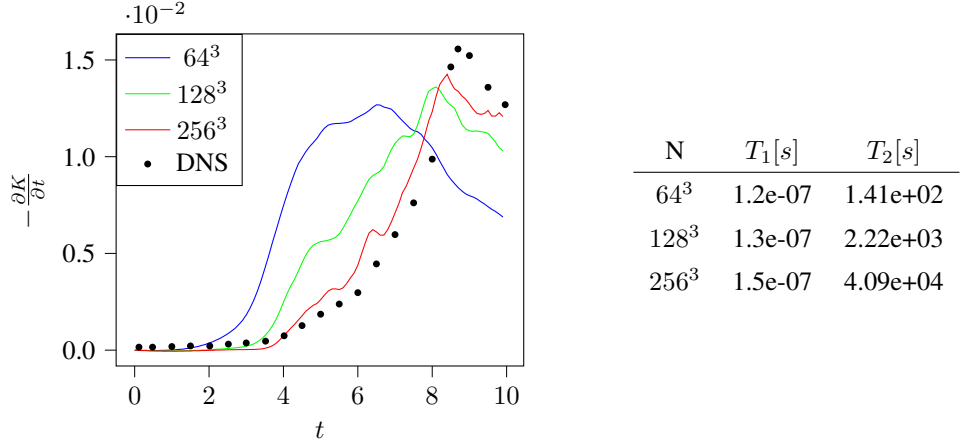

\subsubsection{Taylor-Green Vortex}

The (inviscid) Taylor-Green vortex is one of the simplest cases to investigate the generation of small scales and thus the transition from laminar 
to turbulent flow \cite{Brachet1983,Brachet1984}. In the early stages of the Taylor-Green vortex, the flow field is laminar and strongly anisotropic.
Subsequently, energy is transferred to higher wave numbers and the flow becomes turbulent. In Figure \ref{tgv}, we compare the characteristic growth and decay of the dissipation rate 
for multiple resolutions $N$ to the DNS of Brachet et al. \cite{Brachet1983}. Furthermore, the efficiency of our code for this specific test-case is shown.

\subsubsection{Double Mach reflection}

\begin{figure}[t]
    \centering
    \begin{tikzpicture}
        \node[inner sep=0cm] (a1) at (0.0,0) {\includegraphics[scale=0.25]{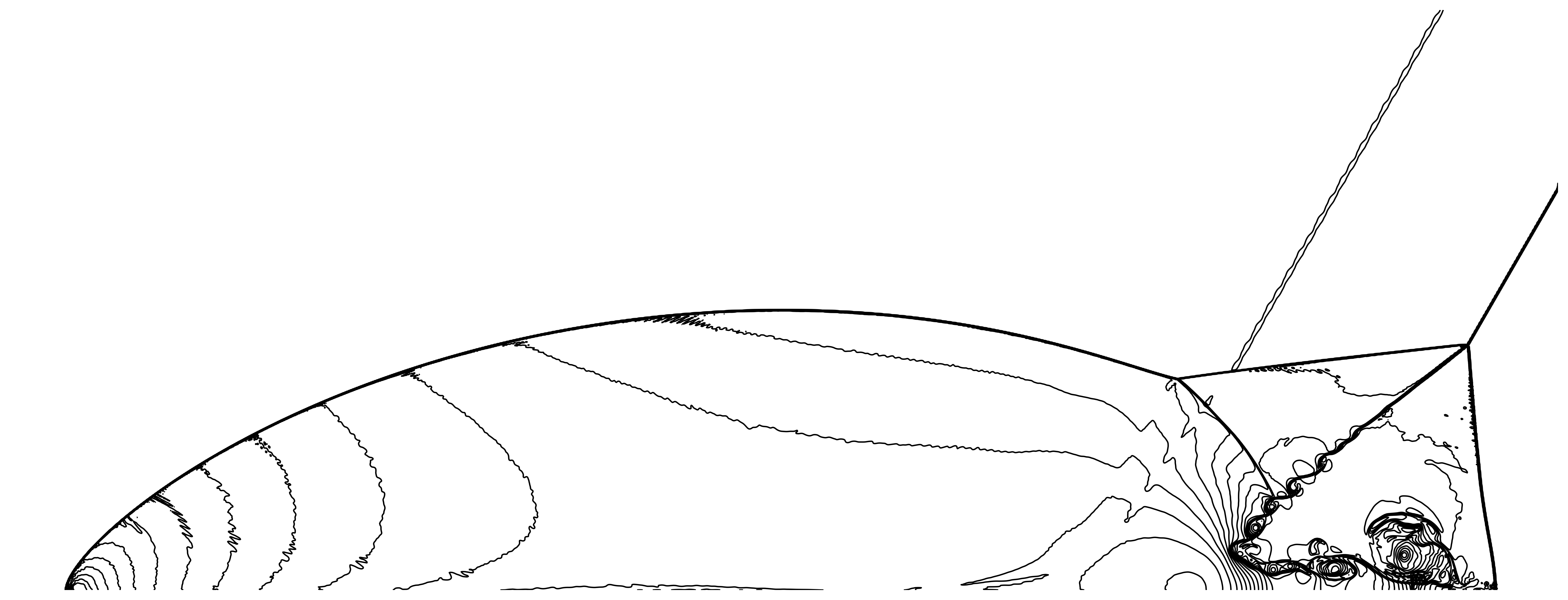}};
    \end{tikzpicture}
    \caption{Instantaneous density isolines of the double Mach reflection.}
    \label{double_mach}
\end{figure}

The Double Mach reflection is a standard test to evaluate the quality of Euler solvers \cite{Kemm2016}.
First suggested by Woodward et al. \cite{Woodward1984}, the case consists of a strong shock hitting a wedge with an inclination
of 30 degrees. While the shock travels up the wedge, a self similar structure evolves. In Figure \ref{double_mach}, this structure is visualized
using the instantaneous density isolines. We simulated this case on a resolution of $2048\times3072$. Note that the actual height of computational domain is 
four times higher than what is shown in Figure \ref{double_mach}, ensuring that the influence of the north boundary is neglectable.

\subsection{Backward-pass}
Automatic differentiation yields the opportunity to optimize and learn numerical schemes from data by end-to-end optimization through a numerical simulator.
As has been done in previous applications, a numerical scheme can be learned by minimizing a loss between a predicted trajectory and a ground truth trajectory.
We optimize the popular local Lax-Friedrichs (LxF) flux function (also known as Rusanov flux function) which is defined by 

\begin{align}
    \bm{F}_{LxF} = \frac{1}{2} (\bm{F}_L + \bm{F}_R) - \frac{1}{2} \alpha (\bm{U}_R - \bm{U}_L).
    \label{eq:Rusanov}
\end{align}

Here, $\bm{U}_{L,R}$ and $\bm{F}_{L,R}$ are the left and right sided cell-face reconstructions of the conservatives and flux, respectively, and
$\alpha$ is the scalar numerical viscosity.
For the classical Rusanov method the numerical viscosity is defined as $\alpha_{Rus} = \max{|u-c|, |u+c|}$, 
where $u$ is the cell-face normal velocity and $c$ is the local speed of sound.
It is well known that although the Rusanov method yields a stable solution without explicitly resolving the interface Riemann problem,
the excess numerical diffusion leads to a smeared out solution.
As a simple demonstration of the AD-capabilities of our compressible fluid solver, we introduce the Rusanov-NN flux 
with the dissipation $\alpha^{NN}_{Rus} = NN \left( \vert \Delta u \vert, u_M, c_M, \vert \Delta s \vert \right)$ to be optimized.
The dissipation is output of a multi-layer perceptron which takes as inputs the jump in normal velocity $\Delta u = \vert u_R - u_L \vert$,
the mean normal velocity $u_M = \frac{1}{2} (u_L + u_R)$, the mean speed of sound $c_M = \frac{1}{2} (c_L + c_R)$, and the entropy jump $\Delta s = \vert s_R - s_L \vert$.

\begin{figure}[t]
    \centering
    \begin{tikzpicture}
        \node[inner sep=0cm] (a1) at (0.0,0) {\includegraphics[scale=0.4]{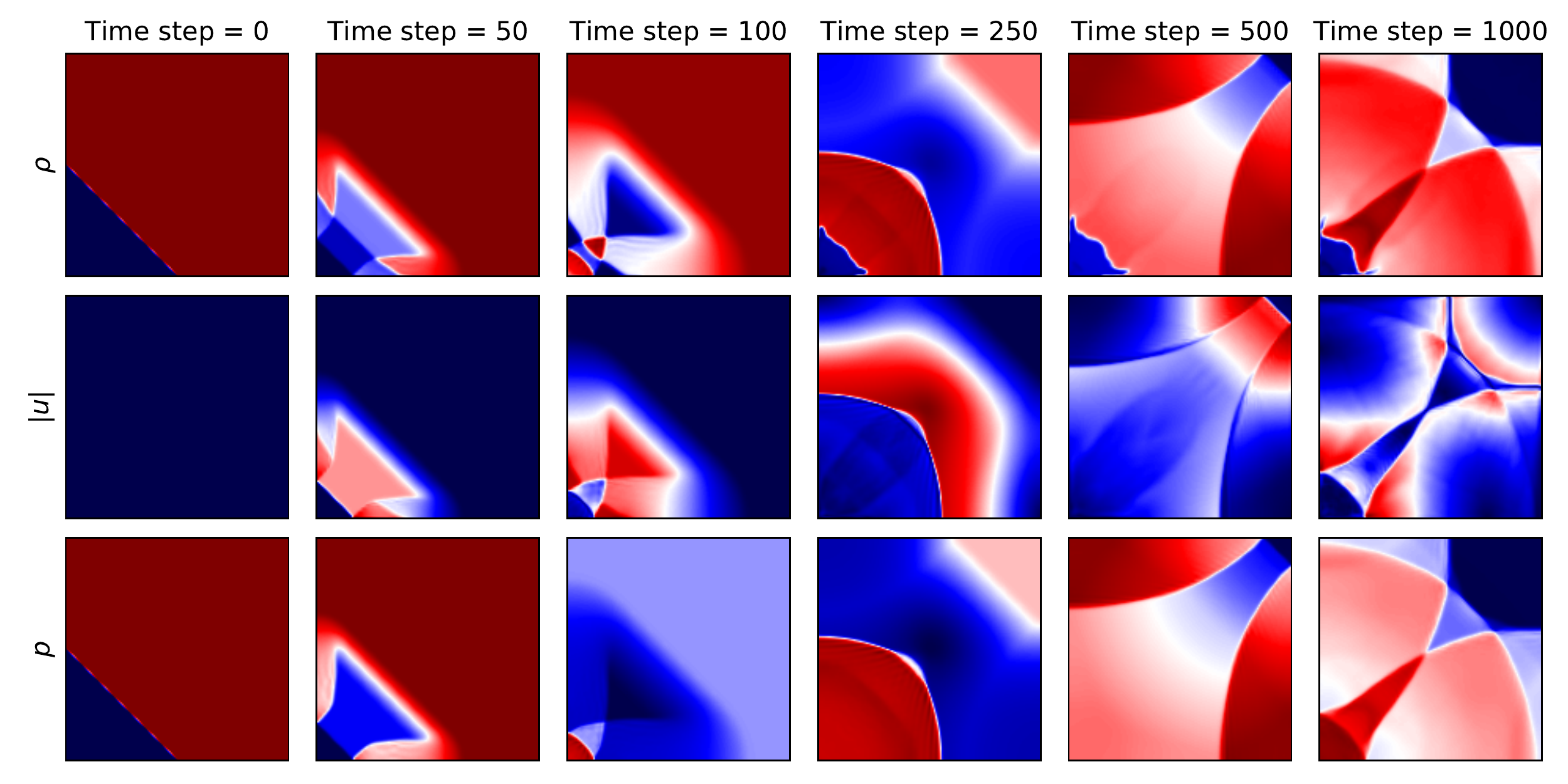}};
    \end{tikzpicture}
    \caption{Training data for density, absolute velocity, and pressure at a resolution of $128 \times 128$.}    
    \label{fig:TrainData}
\end{figure}  

\begin{figure}[t!]
    \centering
    \begin{tikzpicture}
        \node at (-1,0) {\includegraphics[scale=0.4]{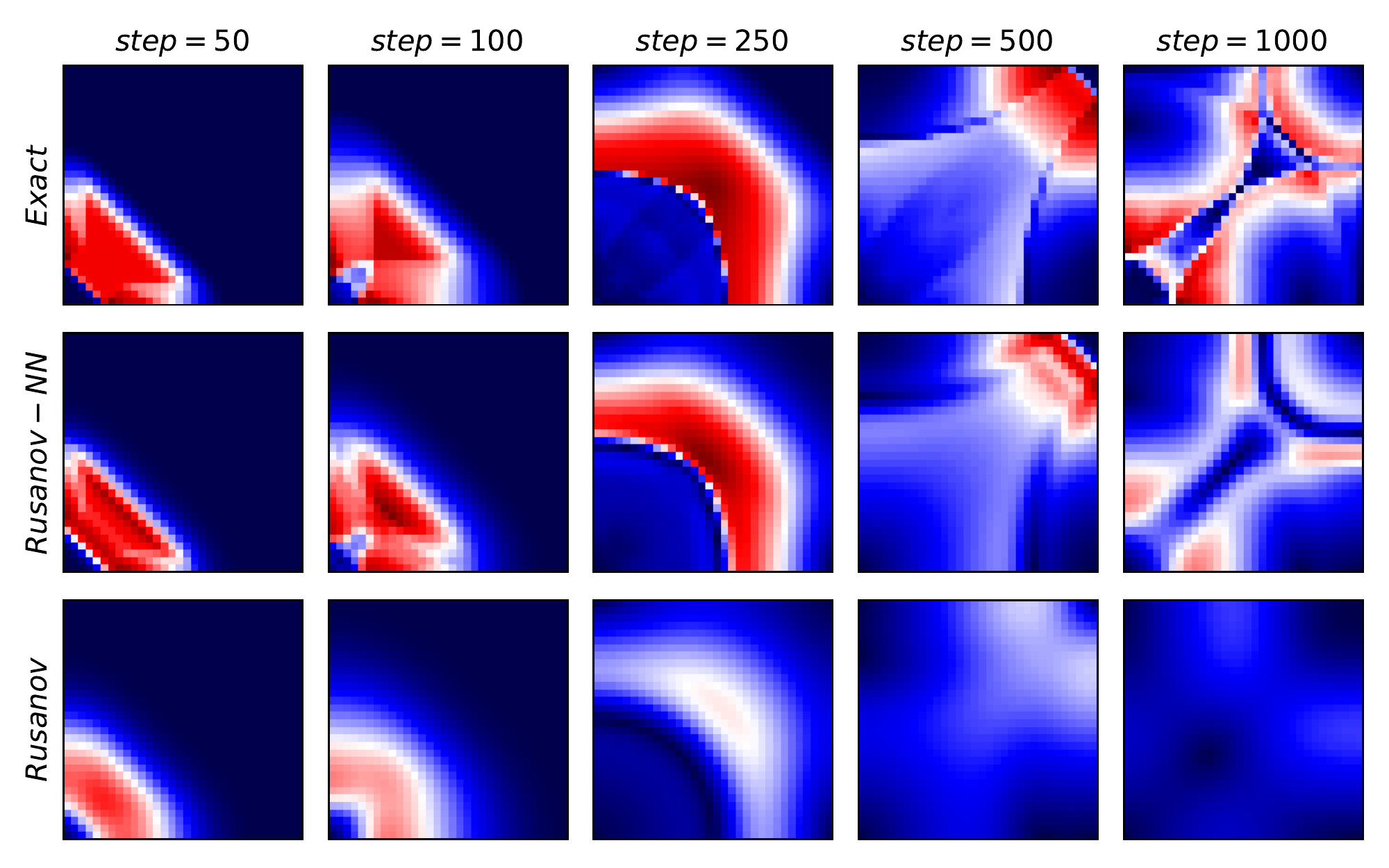}};
        \node at (6,-0.1) {\input{figures/rel_error.tex}};
    \end{tikzpicture}
    \caption{(Left) Absolute velocity fields for the exact (CG $32 \times 32$), ML ($32 \times 32$), and baseline ($32 \times 32$) solvers.
    (Right) Relative $L_2$ percentage error for density (red) and pressure (blue). The gray line indicates the training horizon.}    
    \label{fig:RusanovNN}
\end{figure}

We set up a highly resolved simulation of an implosion testcase to generate the ground trouth trajectory.
The high resolution simulation is run on $128 \times 128$ cells with a WENO5-JS cell-face reconstruction, TVD-RK3 integration scheme, and the HLLC Riemann solver.
The initial and boundary conditions are shown in appendix \ref{IC_FI}.
A trajectory of 3000 time steps is generated. Examplary time snapshots for density, absolute velocity, and pressure are visualized in Fig. \ref{fig:TrainData}.
The left column in Fig. \ref{fig:TrainData} shows the initial condition which is a diagonally placed jump in pressure and density.
A shock and a rarefaction wave are emanated from the initial discontinuity and travel along the diagonal $y = x$.
The shock is propagating towards the lower left corner and is reflected by the walls resulting in a double Mach reflection,
while the rarefaction wave is travelling into the open domain.
We obtain the ground truth data after coarse-graining onto $32\times32$ points, see top row of Fig. \ref{fig:RusanovNN}.

Here, the ML model is trained in a supervised fashion by minimizing a loss between the coarse-grained (CG) high-resolution simulation (labeled as "Exact" or ground-truth) and the simulation produced by the ML model on a coarse grid.
The loss function is defined as the mean-squared error between the predicted and ground truth primitive state vectors $\tilde{\bm{W}} = \left[\rho, u, v, p\right]$ and $\bm{W}$ over a trajectory of length $n_T$, 

\begin{align}
    L = \frac{1}{n_T}\sum_{i = 1}^{n_T} MSE(\bm{W}(t_i), \tilde{\bm{W}}(t_i)).
    \label{eq:Loss}
\end{align}

The training data set consists of the first $2000$ time steps of the coarse grained reference solution. During training, the model is unrolled for
$15$ time steps. 
We use the Adam optimizer with a constant learning rate $5e-4$ and a batch size of $100$.
The Rusanov-NN model is trained for $1000$ epochs.
Figure \ref{fig:RusanovNN} compares the Rusanov and the Rusanov-NN flux function over the full length of a simulation.
Although we have trained on trajectories of length $15$, we evaluate the model for a much longer trajectory of $3000$ time steps.
The NN-Rusanov flux stays stable over the course of the simulation and consistenly outperforms the Rusanov flux.
The ML model even performs very well outside the training set (time steps larger than $2000$).
The NN-Rusanov flux is less dissipative than the classical Rusanov scheme and recovers small scale flow structures very well, see time step 100 in Fig. \ref{fig:RusanovNN}.


\section{Conclusion}
\label{sec:Conclusion}

In this work, a fully-differentiable three-dimensional compressible fluid solver is introduced.
Using state-of-the-art high-order numerical methods, the capabilities and effieciency of the code is demonstrated
on several classical shock-dominated and turbulent fluid flow test cases.
Furthermore, we show that our differentiable solver is capable of end-to-end optimization of neural networks substituting conventional numerical schemes. 
We believe that fully-differentiable CFD algorithms, written in a rapid-prototyping language as the one presented here, 
have enormous potential and will enable new avenues of research.

\begin{ack}
D.A.B. and A.B.B. thank Steffen J. Schmidt for fruitful discussions.\\

This project has received funding from the European Research Council (ERC) under the European Union's Horizon 2020 research and innovation programme (grant agreement No. 667483).
\end{ack}


\appendix
\section{Rayleigh-Taylor instability}
\label{IC_RTI}

The computational domain is $x, y \in [0, 0.25] \times [0, 1]$. The initial condition is given by
\begin{align}
    \left( \rho, u, v, p \right) = \begin{cases}
        \left(2.0, 0.0, -0.025 a \cos(8 \pi x), 1 + 2y\right)   & \text{if}\  y \leq 0.5,\\
        \left(1.0, 0.0, -0.025 a \cos(8 \pi x), 1.5 + y\right)  & \text{if}\  y > 0.5,
    \end{cases}
    \label{eq:RayleighTaylor}
\end{align}
with the speed of sound $ a = \sqrt{\gamma p / \rho} $ and the ratio of specific heats $\gamma = 5 / 3$.
The east and west boundary conditions are symmetry. The north and south boundary conditions are dirichlet imposing zero velocities $u$ and $v$. 

\section{Taylor-Green vortex}
\label{IC_TGV}

The computational domain is $x, y, z \in [0, 2\pi] \times [0, 2\pi] \times [0, 2\pi]$ with periodic boundary conditions for all boundaries.
The initial condition is given by
\begin{align}
    \begin{pmatrix}
        \rho \\
        u \\ 
        v \\
        w \\
        p
    \end{pmatrix}
    =
    \begin{pmatrix}
        1.0 \\
        \sin(x/L)\cos(y/L)\cos(z/L) \\ 
        -\cos(x/L)\sin(y/L)\cos(z/L) \\
        0.0 \\
        \frac{1}{\gamma Ma^2} + \frac{1}{16} (\cos(2x/L)\cos(2y/L)\cos(2z/L))
    \end{pmatrix},
    \label{eq:TaylorGreen}
\end{align}
with $\gamma = 1.4$ and $Ma=0.1$.

\section{Double Mach reflection}
\label{IC_DMR}
The computational domain is  $x, y \in [0, 4.0] \times [0, 6.0]$. The initial condition is given by
\begin{align}
    \left( \rho, u, v, p \right) = \begin{cases}
        \left(\rho_l, u_l, v_l, p_l\right) & \text{if}\ y > \sqrt{3}\left(x-\frac{1}{6}\right),\\
        \left(\rho_r, u_r, v_r, p_r\right) & \text{if}\ y \leq \sqrt{3}\left(x-\frac{1}{6}\right),
    \end{cases}
    \label{eq:DoubleMach}
\end{align}
with 
\begin{equation}
    \begin{pmatrix}
        \rho_l \\
        u_l \\ 
        v_l \\
        p_l
    \end{pmatrix}
    =
    \begin{pmatrix}
        8.0 \\
        7.145 \\ 
        -4.125 \\
        116.5
    \end{pmatrix}, \qquad
    \begin{pmatrix}
        \rho_r \\
        u_r \\ 
        v_r \\
        p_r
    \end{pmatrix}
    =
    \begin{pmatrix}
        1.4 \\
        0.0 \\ 
        0.0 \\
        1.0
    \end{pmatrix}.
\end{equation}
The north and east boundary conditions impose zero gradients. The west boundary condition is dirichlet imposing $\left(\rho_l, u_l, v_l, p_l\right)$.
The south boundary condition is a combination of dirichlet, imposing $\left(\rho_l, u_l, v_l, p_l\right)$, and no slip wall.

\section{Full Implosion}
\label{IC_FI}
The initial conditions are given by 
\begin{align}
    \left( \rho, u, v, p \right) = \begin{cases}
        \left(0.14, 0, 0, 0.125\right)    & \text{if}\ x + y \leq 0.15,\\
        \left(1, 0, 0, 1\right)  & \text{if}\ x + y > 0.15,
    \end{cases}
    \label{eq:FullImplosion}
\end{align}
on the computational domain $x, y \in \left[0, 0.3\right] \times \left[0, 0.3\right]$ with symmetry boundary conditions.

\end{document}